\documentclass[apjl]{emulateapj}
\lefthead{ TSUJIMOTO}
\righthead{Fossil Imprints of Outflow from the Galactic Bulge}

\def\ltsima{$\; \buildrel < \over \sim\;$}
\def\ltsim{\lower.5ex\hbox{\ltsima}}
\def\gtsima{$\; \buildrel > \over\sim \;$}
\def\gtsim{\lower.5ex\hbox{\gtsima}}
\def\ms{$M_{\odot}$ }

\slugcomment{Accepted for publication in ApJ Letters}

\begin{document}
\title{Fossil Imprints of Outflow from the Galactic Bulge in Elemental Abundances of Metal-Rich Disk Stars}

\author{Takuji Tsujimoto}

\affil{National Astronomical Observatory, Mitaka-shi,
Tokyo 181-8588, Japan; taku.tsujimoto@nao.ac.jp}

\begin{abstract}
We explore the elemental abundance features of metal-rich disk stars, highlighting the  comparisons made with those of the recently revealed Galactic bulge stars. A similarity between two of the comparisons leads to a new theoretical picture of the bulge-disk connection in the Galaxy, where a supermassive black hole resides at the center.  We postulate that a metal-rich outflow, triggered by feedback from a black hole, was generated and quenched the star formation, which had lasted several billion years in the bulge. The expelled gas cooled down in the Galactic halo without escaping from the gravitational potential of the Galaxy. The gas gradually started to accrete to the disk around five billion years ago, corresponding to the time of sun's birth, and replaced a low-metallicity halo gas that had been accreting over nearly ten billion years. The metal-rich infalling gas, whose elemental abundance reflects that of metal-rich bulge stars, mixed with the interstellar gas already present in the disk. Stars formed from the mixture compose the metal-rich stellar disk. This scheme is incorporated into models for chemical evolution of the disk. The resultant elemental features are compatible with the observed abundance trends of metal-rich disk stars, including the upturning feature exhibited in some [X/Fe] ratios, whose interpretation was theoretically puzzling. Furthermore, the predicted abundance distribution function of disk stars covers all observational facts to be considered: (i) the deficiency of metal-poor stars, (ii) the location of peak, and (iii) the extended metal-rich tail up to [Fe/H]$\sim +0.4$.  
 \end{abstract}


\section{Introduction}

Feedback from supermassive black holes at the innermost regions of galaxies has a crucial influence on the evolution of their host galaxies \citep{Silk_98}.  While the growth of black holes through gas accretion is regulated by its associated feedback, the feedback also influences the host on a large scale through energy input into the gas surrounding the black holes,  which halts star formation and induces a powerful outflow \citep[e.g.,][]{Springel_05}. Recent theoretical works \citep[e.g.,][]{Matteo_05,Schawinski_06} suggest that the processes related to feedback lead to the observed relationship between the black hole mass and the velocity dispersion of the galaxy spheroidal component \citep{Magorrian_98,Ferrarese_00,Gebhardt_00}. 

It is evident that the Galaxy possesses a supermassive black hole, with a mass of $\sim 3\times 10^6$\ms at the Galaxy center \citep{Genzel_97,Schodel_03}. Because of the short distance, the Galaxy is the best-suited laboratory to probe the past of other galaxies hosting supermassive black holes and that  have been affected by their feedback. Now is the time to pursue this subject using a powerful tool to trace the mechanisms of galaxy formation, e.g., elemental abundances of stars, especially since we have obtained detailed elemental abundances of individual stars in the Galaxy, including member stars belonging to the bulge.

The Galactic bulge is old \citep[e.g.,][]{Ortolani_95}. No clear evidence of young stellar population can be detected in the color-magnitude diagram of the Galactic bulge stars \citep{Feltzing_00}, although elemental abundances of bulge stars  \citep{Fulbright_07} with a signature of Type Ia supernovae (SNe Ia) imply that the period of star formation should be in the order of a billion years. In any event, the process of expelling the gas from the bulge to cease star formation at a young age of the Galaxy is required at the final stage of bulge formation, which can be identified with outflow energized by any past activity in the existing supermassive black hole. 

Outflow from the Galactic bulge should be well enriched because its elemental feature is the end result of chemical evolution in the Galactic bulge. Thus, the metallicity of outflow will be almost equivalent to the highest among the bulge stars, whose metallicities range from $-1.3\leq$[Fe/H]$\leq +0.5$ \citep{Fulbright_06}. Such a metal-rich outflow crucially influences the chemical evolution of the Galaxy because of its high metallicity, as long as the gas released from the Galactic bulge does not have sufficient energy to escape from the gravitational potential of the Galaxy. This may be the likely fate of any outflow originating from a supermassive black hole with a relatively small mass like ours. 

Recently, detailed elemental abundances of the Galactic bulge stars have been revealed \citep{Fulbright_07}.  One of the most remarkable abundance features is the enhancement of [$\alpha$-element/Fe] in comparison with disk stars in the overall metallicity range, including metal-rich stars above solar metallicity. These characteristics can be used for chemical tagging of an outflow. Considering the cooling process that acted on the outflow while moving in the Galactic halo, a promising site where we could search for this tag would possibly be the Galactic disk, although the process to gain angular momentum should be validated by future numerical simulations.

Disk stars\footnote{In this {\it letter}, disk stars include only stars belonging to the thin disk component in the Galaxy.} present a puzzling problem associated with their elemental feature for [Fe/H]$\gtsim 0$. It is well understood that the observed decreasing [$\alpha$-element/Fe] trends with an increasing [Fe/H] for the range of  $-1\ltsim$[Fe/H]$\ltsim 0$ are caused by a delayed Fe supply from SNe Ia compared with Type II SNe (SNe II). On the contrary, a theoretical interpretation of their abundance trends for [Fe/H] $\gtsim 0$ has not yet been obtained. Strangely, the [$\alpha$-element/Fe] ratios, except for [O/Fe], start to upturn from [Fe/H]$\sim 0$ and continue to rise until reaching the highest metallicity [Fe/H]$\sim +0.4$ \citep{Reddy_03,Bensby_05}. Such upturns can also be seen in other elements, such as odd elements (Na, Al), and some iron-group elements (Ni, Zn). It is incorrect to attribute these upturns to metallicity-dependent SN II yields, especially because this feature is observed with both even- and odd-numberd elements. Alternatively, some new insights from the viewpoint of galaxy formation should be introduced to understand the late stage of chemical evolution in the Galactic disk imprinted in elemental abundances of metal-rich disk stars. 

Models of chemical evolution require that the Galactic disk is formed by a continuous infall of gas from the halo, to solve the so-called G-dwarf problem \citep[e.g.,][]{Pagel_97}.  Conventionally, it is assumed that an infalling gas has very low-metallicity over the entire period of disk formation. In practice, models adopting this assumption apparently reproduce the observed abundance distribution function (ADF) of disk stars, as attempted by many authors \citep[e.g.,][]{Yoshii_96, Chiappini_97}. However, as will be discussed in the next section, there exist a fatal problem with the present models when trying to reproduce the overall ADF by paying attention to not only its metal-poor tail but also its metal-rich side.

Here, we first propose that there exists a switchover from accretion of low-metallicity halo gas to that of metal-rich gas onto the disk at the time when the metallicity in the disk nearly reaches solar metallicity. This is followed by the scenario that a metal-rich gas blown away from the Galactic bulge cooled down radiatively in the halo and finally settled down on the disk that had been formed from a low-metallicity infall up until that time.  Based on this theoretical scheme, the model of chemical evolution is designed and found to  successfully reproduce the observed abundance trend of disk stars in the range $-1\leq$[Fe/H]$\leq +0.4$. Our scenario is supported by the presence of stars with super solar metallicities such as  [Fe/H]$= +0.2-0.4$. We will begin with a discussion on this matter.
 
\section{Implications from Abundance Distribution Function of Disk Stars}

The distinct features of the ADF of solar neighborhood disk stars, confirmed by many studies thus far \citep[e.g.,][]{Wyse_95,Rocha-Pinto_96,Nordstrom_04} are (i) the deficiency of metal-poor stars: there are few stars below [Fe/H]$\ltsim -1$, i.e., the G-dwarf problem, and (ii) the peak located at [Fe/H]$_{\rm peak}$$\sim -0.2 - -0.1$. In addition to these observational facts, spectroscopic observations of elemental abundances for  metal-rich disk stars \citep{Feltzing_98,Bensby_05} have revealed that chemical enrichment in the solar neighborhood has continuously proceeded until [Fe/H]$\sim +0.4$. This information, relating to how far the stellar metallicity extends to a metal-rich direction, should be the third ingredient to be highlighted in the reproduction of the observed ADF.  

\begin{figure}[t]
\vspace{0.2cm}
\begin{center}
\includegraphics[width=7cm,clip=true]{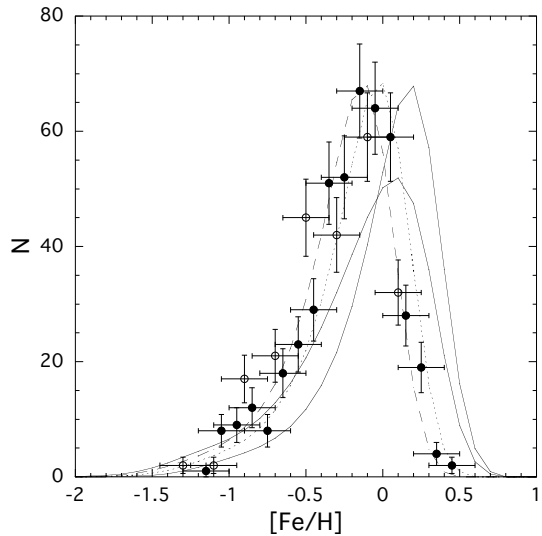}
\end{center}
\vspace{0.3cm}
\caption{Abundance distribution function of disk stars against the iron abundance. Solid curves are the predictions of the model with [Fe/H]$_{\rm present}=+0.4$ and $t_{\rm Ia}$=1.5 Gyr for $t_{\rm in}$=3 Gyr (left) and $t_{\rm in}$=5 Gyr (right), respectively. Results of the model with $t_{\rm Ia}$=1.5 Gyr and $t_{\rm in}$=5 Gyr for  [Fe/H]$_{\rm present}=+0.03$ and  $+0.15$ are shown by dashed curve and dotted curve, respectively. These calculated distributions are convolved using the Gaussian with a dispersion of 0.15 dex in [Fe/H]. Filled and open circles represent data taken from \citet{Edvardsson_93} and \citet{Wyse_95}, respectively.  The model distributions and the observed one by \citet{Wyse_95} are normalized to coincide with the total number of the sample stars used by \citet{Edvardsson_93}.}
\end{figure}

Next, we will try to reproduce the ADF of disk stars, paying special attention to the metal-rich side, which has not been fully considered by the previous studies. For this purpose, the procedure to solve the equations for chemical evolution adopted by \citet{Yoshii_96} is used, because in this approach, the present metallicity [Fe/H]$_{T_{\rm G}}$ of gas, which corresponds to the highest metallicity of disk stars, is not an outcome but an input parameter. It means that the value of  [Fe/H]$_{T_{\rm G}}$ can be chosen arbitrarily. This, together with the present gas fraction $f_g (T_{\rm G})$, is combined with four theoretical inputs composed of (i) the power index $n$ of the star formation rate (SFR) in Schmidt law, (ii) the timescale $t_{\rm in}$ of infall, (iii) the lifetime $t_{\rm Ia}$ of SN Ia progenitors, and (iv) the metallicity [Fe/H]$_{\rm infall}$ of the infalling gas, to determine the chemical evolution of the disk. For (i), it is assumed that the SFR is proportional to the gas density, i.e., $n=1$. For (iv), an infall from the halo under consideration implies a very low metallicity, likely to be, [Fe/H]$_{\rm infall}$$\ll -1$ reflecting the metallicity of halo stars. Since this value is irrelevant to the prediction of ADF, as long as  [Fe/H]$_{\rm infall}$$<-1$ \citep{Yoshii_96}, we assume  [Fe/H]$_{\rm infall}=0$. In the end, each combination of $t_{\rm in}$ and $t_{\rm Ia}$ yields various forms of the ADF.

The solid curves in Fig.~1 show the resultant ADFs with the choices of $t_{\rm Ia}$=1.5 Gyr and $t_{\rm in}$=3, 5 Gyr, together with [Fe/H]$_{T_{\rm G}}$=+0.4 and $f_g (T_{\rm G})$=0.15 \citep{Young_91}. Here [Fe/H]$_{T_{\rm G}}$ is assumed to equal [Fe/H] of the most metal-rich local F/G stars \citep{Bensby_05}. The age  $T_{\rm G}$ of a galaxy is set to be 14 Gyr. Note that we convolve the calculated ADF using the Gaussian with a dispersion of 0.15 dex in [Fe/H], which mimics the error of the data. Obviously, the predicted ADFs are biased to metal-rich as compared with the observations. Any combinations of the four input parameters, together with an allowable uncertainty in the observed $f_g (T_{\rm G})$, are unable to avoid positioning the peak beyond solar metallicity as long as the observed value for [Fe/H]$_{T_{\rm G}}$ is adopted. We performed another calculation with the assumption of [Fe/H]$_{\rm G}$=+0.15. This value corresponds to the mean metallicity at present in  the observed age-metallicty relation of nearby stars \citep{Nordstrom_04, Reid_07}. The resultant ADF is shown to be still shifted to metal-richer than the observation (dotted curve). On the other hand, if we assume a smaller [Fe/H]$_{T_{\rm G}}$, such as +0.03 (dashed curve), the calculated ADF matches the observed one, as already shown in the previous studies \citep[e.g.,][]{Yoshii_96}. 

In summary, the simultaneous reproduction of both [Fe/H]$_{T_{\rm G}}$$\geq +0.15$ and [Fe/H]$_{\rm peak} <0$ is hard to realize through the conventional scheme adopted for chemical evolution of the Galactic disk. Thus, some modification of the constructed model is required.

\section{Abundance Trends of Metal-Rich Disk Stars}

In this section, we will examine the abundance trends of metal-rich disk stars with [Fe/H]$\gtsim 0$ to persuade readers that the key factor required to solve the problem in the existing models is tightly linked to the activity and formation of the Galactic bulge. Figure 2 shows the comparison between abundance features in the [X/Fe] vs.~[Fe/H] diagram for six elements of disk stars and metal-rich bulge stars. As already mentioned in Section 1, the characteristic feature of $\alpha$-elements, such as Mg, Si, Ca, and Ti, in comparison with Fe for metal-rich disk stars, is the upturn with an increase in [Fe/H]. Two upper panels give a sample of $X$=Mg and Si, demonstrating that this feature is surely observed in $0 \ltsim$[Fe/H]$\ltsim +0.4$. Although the origin of these upturn has been a totally open question, superposition of the observed data for metal-rich bulge stars on those of the disk stars leads to a possible indication for solving the puzzling upturn. Our proposed view of abundance trends between two different populations is that the [Mg (Si)/Fe] ratio of disk stars gradually {\it approaches} that of the metal-rich bulge stars in relation to the increase in [Fe/H]. This interpretation also holds true in the cases of [O/Fe] and [Mn/Fe],  where the abundance trend in a metal-rich regime inherits that for [Fe/H]$\ltsim 0$ (middle panels of Fig.~2). In addition, odd elements such as Na and Al, which show an upturn in their ratios to Fe like the case of $\alpha$-elements, also have enhanced ratios of bulge stars ahead in the direction of their upturn trends  (lower panels of Fig.~2). 

\begin{figure}[t]
\vspace{0.2cm}
\begin{center}
\includegraphics[width=7cm,clip=true]{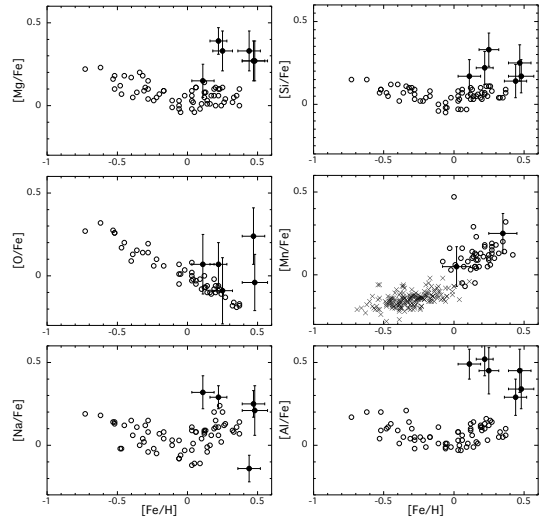}
\end{center}
\vspace{0.3cm}
\caption{Correlations of [X/Fe] with [Fe/H] for disk stars (open circles and crosses) together with metal-rich ([Fe/H]$> 0$) bulge stars (filled circles). Except for Mn, filled and open circles represent data from \citet{Fulbright_07} and \citet{Bensby_05}, respectively. The Mn abundances are taken from \citet{McWilliam_03}(filled circles), \citet{Feltzing_98}(open circles), and \citet{Reddy_03}(crosses).}
\end{figure}
 
However, this poses the question of which theoretical scheme can be applied to the potential bulge-disk connection to understand their stellar elemental relationship. A possible answer is that metal-rich disk stars are formed from a mixture of an infalling metal-rich gas originally ejected from the bulge and the remaining gas of a star formation in the disk. In this case, their abundance pattern is determined by the combination of heavy elements ejected from the bulge at the end of star formation, along with those elements that are already present in the interstellar gas of the disk. In accordance with an accumulation of infall with time, the source of heavy elements in stellar abundances is gradually dominated by those in a metal-rich infalling gas. This results in an abundance trend directed to the elemental ratios of metal-rich bulge stars. Furthermore, an iron contained in a metal-rich infalling gas will promote an increase in the [Fe/H] of stars, which should result in an extended metal-rich tail of ADF.
   
\section{Metal-Rich Outflow from the Galactic Bulge}

\begin{figure}[t]
\vspace{0.2cm}
\begin{center}
\includegraphics[width=7cm,clip=true]{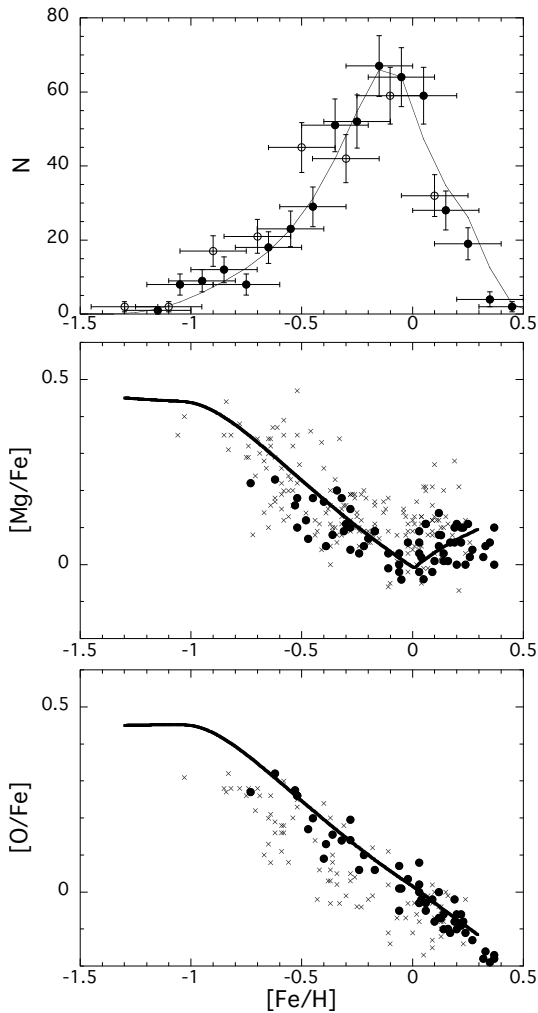}
\end{center}
\vspace{0.3cm}
\caption{Features of chemical evolution in the Galactic disk predicted by the model in which a metal-rich infall whose elemental abundances match those of metal-rich bulge stars are introduced in the late stage of disk formation. See the text for details. The observational data in the upper panel is same as in Fig.~1. In the lower two panels, the filled circles and crosses are taken from \citet{Bensby_05} and \citet{Edvardsson_93}, respectively.}
\end{figure}

By incorporation of a switchover of the origin of an infalling gas from the halo to the bulge into the model, we describe the evolution of stellar abundances in the Galactic disk. The basic picture of the model is that the disk was formed through an infall of material from outside the disk region, and that in the first $9\times10^9$yrs, the halo provided a very low metallicity gas for the disk; whereas, in the subsequent period, a metal-rich gas originating from the bulge replaced the former halo gas. The validity of such a longtime delay of metal-rich infall should be investigated by numerical simulations that will clarify how 
the outflow behaves in the halo under physical processes such as  a radiative cooling, interactions with halo substructure, and a possible heating from a prolonged star formation in the thick disk \citep{Bensby_07}.  For simplicity, we propose a formula that is proportional to $t\exp(- t/t_{\rm in})$ for the whole duration. In this formula, the timecsale and metallicity of an infalling gas is assumed to be $t_{\rm in}$=5 Gyr and [Fe/H]$_{\rm in}$=$-1.3$ for 0$\leq t{\rm (Gyr)} \leq$9 and $t_{\rm in}$=3 Gyr and [Fe/H]$_{\rm in}$=$+0.2$ for 9$< t{\rm (Gyr)} \leq$14. Each infall rate is normalized so that an integration over $t=0$ to $\infty$ is an unit.  The value of $t_{\rm in}$ for the latter is adjusted to give the mass ratio of 4:1 between the former metal-poor gas and metal-rich one, which is roughly equivalent to the ratio of metal-poor to metal-rich stars in the solar neighborhood. We adopt a somewhat smaller [Fe/H]$_{\rm in}$ for the bulge origin, taking into account the dilution by low-metallicity  gas during its passage through the halo. Abundances for other heavy elements follow those of observed halo stars or metal-rich bulge stars for each infall (see below).  

Here, we adopt the usual procedure of solving the basic equations for chemical evolution, in which the SFR and the stellar initial mass function (IMF) are given as input parameters. The SFR is assumed to be proportional to $f_g$ with a constant rate coefficient of 0.4 Gyr$^{-1}$. For the IMF, we assume a power-law mass spectrum with a Salpeter slope of $1.35$, which is combined with the nucleosynthesis yields of SNe Ia and II taken from \citet{Tsujimoto_95}. For the lifetime of SN Ia progenitors, $t_{\rm Ia}$=1.5 Gyr is adopted.

The results are shown in the three panels of Fig.~3. The top panel shows the calculated ADF compared with the observations. As shown in the lower two panels, the metallicity is enriched up to [Fe/H]$\sim +0.3$. Despite the attainment of super solar metallicity, the predicted ADF is not skewed to metal-rich,  unlike the models presented in Fig.~1. The success in reproducing the right position of the peak, as well as the precise features of both sides of its peak, is attributable to the combination of a low-metallicty infall with a prolonged timescale at the early stage of disk formation, plus the subsequent metal-rich infall. 
The middle panel shows the predicted evolutionary change in [Mg/Fe] with [Fe/H] for disk stars. Here, we assume that the [Mg/Fe]$_{\rm infall}$ ratio of an infalling gas is $+0.45$ for the halo origin and $+0.3$ for the bulge origin. First, a delayed supply of iron with a negligible amount of Mg from SNe Ia producs a decreasing trend of [Mg/Fe] for [Fe/H]$\ltsim 0$ from a low-metallicty infalling gas. Then, it results in [Mg/Fe]$\sim 0$ at [Fe/H]$\sim 0$ for a timescale of about ten billion years. Subsequently, high abundances of a metal-rich infall, such as [Fe/H]$_{\rm infall}$=$+0.2$ and [Mg/H]$_{\rm infall}$=$+0.5$, start to influence the stellar abundance patterns so that they continue the direction of evolving [Mg/Fe] upward,  along with the progress of  chemical enrichment. 

On the other hand,  the lower panel shows that the  behavior of the [O/Fe] ratio is different. The [O/Fe]$_{\rm infall}$ ratio of metal-rich infall is expected to be nearly the solar ratio, as exhibited by metal-rich bulge  stars (left middle panel of Fig.~2). As a result, such a  lower  [O/Fe]$_{\rm infall}$ prevents the [O/Fe] trend from turning upward,  and instead it leads to a continuation of the decrease in  [O/Fe] for [Fe/H]$\gtsim 0$, differing from the common upturn feature observed with other $\alpha$-elements. A good agreement with the observation is achieved by the adoption of [O/Fe]$_{\rm infall}$=$-0.2$, which is somewhat lower than the expected ratio. Thus, we predict that larger data sets should find the most metal-rich bulge star have lower [O/Fe] such as $\sim -0.2$.

Finally, we will check whether the supply of metal-rich gas satisfies the amount of metal-rick disk stars.
Since we have neither information on how the metal-rich gas is distributed over the disk nor the ADF in the disk other than the solar neighborhood, the disk should be considered to be one-zone represented by the solar circle. Let's take the mass of a bulge as 25\% of the mass $M_{\rm G}$ of the Galaxy. The mass fraction $\mu$ of gas after the star formation in the bulge can be deduced from the equation $Z=p \ln \mu^{-1}$ ($Z$: metallicity, $p$: the effective yield) as long as a simple closed-box model is applied. 
Given $p$ derived from chemical evolution in the solar neighborhood \citep{Tsujimoto_97}, the metallicity of the bulge implies $\mu\sim 0.1$. This result combined with the reduction in the metallicity of outflow by $\sim 0.2$ dex through mixing with metal-poor halo gas results in $\sim 0.05 M_{\rm G}$ for the mass of metal-rich infalling gas. On the other hand, the mass $0.75 M_{\rm G}$ of disk (gas and star) and the fraction $\sim 20 \%$ of metal-rich ([Fe/H]$>0$) stars give the mass $\sim 0.12M_{\rm G}$ of metal-rich stars. Taking into account that these stars are made of a metal-rich infall mixed with the gas already present in the disk at $t=9$ Gyr and these contributions are nearly equal in total, the mass of metal-rich infalling gas required to make metal-rich disk stars is roughly equivalent to the former estimate.

\section{Conclusions and Discussion}

We propose that the observational facts of (i) elemental abundance features of metal-rich disk stars characterized by an upturn seen for some elements, and (ii) the extended metal-rich tail up to super solar metallicity in the abundance distribution function of disk stars are evidence that an infall outside the disk region is dominated by the gas ejected by the bulge for the recent period of five billion years. A grand scheme for the formation of bulge and disk in the Galaxy based on this context is described as follows. A prolonged star formation proceeds in the bulge, as implied by a clear signature of SNe Ia in the abundances of bulge stars. Then, feedback from the supermassive black hole that resides in the Galactic center terminates star formation in the bulge and triggers a highly-enriched outflow. Owing to the relatively small mass of the black hole, this outflow is not energetic enough to escape from the gravitational potential of the Galaxy. Instead, it spends several billion years in the halo and then cools down radiatively.  Simultaneously, the disk is gradually formed through an accretion of material with very low metallicity from the halo over nearly ten billion years. Then, at the time when the metallicity becomes nearly solar, the cooled gas originally ejected from the bulge starts to infall onto the disk, replacing the former low-metallicty gas. This metal-rich gas accelerates the rate of increase in metallicity and generates super metal-rich stars whose elemental features mimic those of metal-rich bulge stars. It should be, however, noted that a definitive picture of the bulge-disk connection  in the Galaxy should await future numerical simulations, especially to solve the critical issues of outflow on angular momentum, timescale, and the mass distribution of settlement on the disk.

The discussion of whether it is possible to detect a metal-rich infall cannnot be furthered because predicting its present rate is beyond the capability of our simplified model. Even so, if we search hydrogen clouds already detected in the halo for the potential candidate of a metal-rich gas enriched in the bulge, intermediate-velocity clouds could be promising because they appear to be metal-rich ($\sim$ solar) \citep{Richter_01}; on the other hand, high-velocity clouds, such as  Complex C types, are likely to be of low metallicity \citep{Wakker_99, Collins_07}. 

\acknowledgements
The author is appreciative of the anonymous referee for useful comments that helped improve this Letter, and also thanks K. Wada and M. Mori for helpful discussions.

\end{document}